# A Simple Engineering Technique to Estimate the First Derivative of an Empirical Function

*Emmanuil Beygelzimer[1], Yan Beygelzimer[2]*
[1]*OMD-Engineering LLC, Dnipro, Ukraine*
[2]*Donetsk Institute for Physics and Engineering named after A.A. Galkin, National Academy of Sciences of Ukraine, Kyiv, Ukraine*

*Corresponding author: emmanuilomd@gmail.com, tel.: +380 (50) 368-63-42, 49000, Yevropeiska Str. 5 b, Dnipro, Ukraine.

**ABSTRACT**

A heuristic formula for 5-point approximation of the first derivative of an unknown function whose values are measured with an error at unequally spaced points is proposed. The derivative at a given point is calculated using the effective increments of the function and the argument, taking into account the different weight coefficients for near and far measurement points. Simulation modeling on test functions with known derivatives is applied to determine rational values of weight coefficients. Simulation results are described in detail on two test functions, one of which simulates the process of water cooling of a hot steel sheet, the second is a complex oscillatory process with variable frequency and amplitude. It was found that the optimal values of weight coefficients remain approximately the same for essentially different functions, which allows us to recommend the same formula for all cases. In contrast to the classical methods of numerical differentiation of functions tabulated at unequally spaced nodes, the proposed formula simultaneously takes into account the smoothing of empirical data. It is shown that this significantly increases the accuracy of the numerical derivative estimate even in cases where the random error of the function measurement is a very small value, from ± 1%. The formula obtained is recommended for use in solving any problems requiring the estimation of the derivative of an empirical function, including the calculation of the stress-strain state of metal, the description of thermal processes, and the determination of the properties of materials.

## INTRODUCTION

In the practice of engineering research it is often necessary to estimate the first derivative of some unknown function, which is tabulated in non-uniformly distributed empirical points. For example, such a problem arises in determining the stress-strain state of metal in pressure treatment processes [1; 2], in evaluating the heat flow on the basis of temperature values measured at different moments of time [3-6], in calculating differential indices of material properties after digitizing graphs of corresponding integral characteristics measured experimentally [7-9]. One of two approaches is usually used to solve such problems.

The first approach is to approximate the empirical data by an analytical function and then calculate the derivative of this function using analytical differentiation methods. In many cases, however, it is impossible to satisfactorily describe empirical data with an analytical function, at least over a wide interval of variation in the argument. Therefore, the above approach to finding the derivative often cannot be implemented in practice.

The second approach is to find the finite-difference derivative using numerical differentiation methods [10; 11, p. 433, 450]. One of the most common methods of numerical differentiation at unequally spaced nodes is the use of interpolation Lagrange polynomials. For example, when estimating the first derivative of a function by three points, the following expression is used:

$$(y'_i)_{3uL} = y_{i-1} \frac{X_{i,i+1}}{X_{i-1,i} X_{i-1,i+1}} + y_i \frac{X_{i,i-1} + X_{i,i+1}}{X_{i,i-1} X_{i,i+1}} + y_{i+1} \frac{X_{i,i-1}}{X_{i+1,i-1} X_{i+1,i}} \quad (1)$$

where $y'_i$ is the estimation of the first derivative of the function under study ($y$) at the $i$-th point of the data set; $X_{j,k}$ is the interval difference of the argument:

$$X_{j,k} = x_j - x_k \quad (2)$$

$x$ and $y$ with subscripts are the measured values of the argument and function at the corresponding points (for example, $X_{i,i+1} = x_i - x_{i+1}$).

The subscripts in the notation of the first derivative on the left side indicate that this derivative was estimated from three (3) unequally spaced (*u*) points using Lagrange interpolation polynomials (*L*).

More than three empirical points are used to refine the result. For example, the formula for a 5-point numerical estimate of the first derivative in the case of unequal intervals between points is as follows [12]:

$$(y'_i)_{5uL} = D_{i-2} + D_{i-1} + D_i + D_{i+1} + D_{i+2} \qquad (3)$$

where

$$D_{i-2} = y_{i-2} \frac{X_{i,i-1} X_{i+1,i} X_{i+2,i}}{X_{i-2,i-1} X_{i-2,i} X_{i-2,i+1} X_{i-2,i+2}} \qquad (4)$$

$$D_{i-1} = y_{i-1} \frac{X_{i,i-2} X_{i+1,i} X_{i+2,i}}{X_{i-1,i-2} X_{i-1,i} X_{i-1,i+1} X_{i-1,i+2}} \qquad (5)$$

$$D_i = y_i \frac{(X_{i,i-1} + X_{i,i-2}) X_{i+1,i} X_{i+2,i} - X_{i,i-2} X_{i,i-1} (X_{i+1,i} + X_{i+2,i})}{X_{i,i-2} X_{i,i-1} X_{i,i+1} X_{i,i+2}} \qquad (6)$$

$$D_{i+1} = y_{i+1} \frac{X_{i,i-2} X_{i-1,i} X_{i+2,i}}{X_{i+1,i-2} X_{i+1,i-1} X_{i+1,i} X_{i+1,i+2}} \qquad (7)$$

$$D_{i+2} = y_{i+2} \frac{X_{i,i-2} X_{i-1,i} X_{i+1,i}}{X_{i+2,i-2} X_{i+2,i-1} X_{i+2,i} X_{i+2,i+1}} \qquad (8)$$

the other notations are the same as in formulas (1) and (2).

The above finite-difference formulas for calculating $(y'_i)_{3uL}$ and $(y'_i)_{5uL}$ in the case of empirical data analysis have one fundamental drawback. It lies in the fact that the derivative is calculated using polynomials which pass through tabulated values of the function. It is well known that any experimentally found function always has some error in its values, which may very strongly affect the value of its derivative. For this reason, numerical differentiation should be associated with smoothing of experimental data. In [13, p. 327] formulas for differentiation of empirically determined functions with simultaneous smoothing are given, but they are suitable only for equidistant values of the argument and for rather smoothly varying function (the second derivative of which is close to zero).

The **goal** of this paper is to find an effective way to estimate the derivative of an empirical function under two circumstances: 1) unequal intervals between measuring points of the function and 2) the error of the measured values of the function themselves.

## METHODS

As such an effective way, the authors propose a simple formula for the 5-point approximation of the first derivative of a function at the *i*-point:

$$(y'_i)_{5uS} = \frac{Y_{i+2,i-2} + nY_{i+1,i-1}}{X_{i+2,i-2} + nX_{i+1,i-1}} \qquad (9)$$

where $Y_{j,k}$ is the interval difference of the measured values of *y*:

$$Y_{j,k} = y_j - y_k \qquad (10)$$

e.g., $Y_{i+2,i-2} = y_{i+2} - y_{i-2}$; *n* is the weighting factor determining the degree of influence of the nearest points ($x_{i-1}$ and $x_{i+1}$) relative to the farthest ($x_{i-2}$ and $x_{i+2}$). The subscripts 5*uS* in the derivative notation on the left side of expression (9) mean that the estimation is performed on five (5) unequally spaced (*u*) points by the simple (*S* from "Simple") formula.

Formula (9) has no strict justification and is heuristic. Expressions in its numerator and denominator represent some effective values of increment of function and increment of its argument. The coefficient n takes into account different "weights" of increments associated with different distances of points from the point where the derivative is calculated. It will be shown below that for sufficiently different functions, the same value of n gives good results. To determine the rational values of the weight coefficient n, the authors applied simulation modeling on test functions with known derivatives.

The following function is used as one of these test functions:

$$y(x) = a_1 \text{arcctg}(x - a_2) \tag{11}$$

with the theoretical first derivative:

$$y'(x) \equiv \frac{dy(x)}{dx} = -\frac{a_1}{1 + (x - a_2)^2} \tag{12}$$

where $a_1$ and $a_2$ are constant parameters.

So, let us assume that we experimentally investigate some physical process, which actually obeys the law of the test function (11). For example, it may be a process of water cooling of steel sheet after heating or rolling (see, for example, [14]). In this case *x* can denote the time [s] and *y* the temperature [°C]. The function (11) itself is unknown, and we estimate it by measuring in experiment the values of *y* at some discrete values of *x*. The question is, how accurately can we estimate the derivative (12) (in our example, this derivative means the cooling rate) from formula (9) and what value of the weighting factor *n* will provide the most accurate fit?

When answering this question, let's take into account that temperature measurements are performed with some errors. To take this into account, in the simulation each "measured" value is related to the true temperature by the following relationship:

$$y_i = y(x_i)(1 + 2(\delta_i - 0.5)\varepsilon) \tag{13}$$

where $y_i$ is the result of measurements of the unknown temperature *y*(*x*) at the *i*-th time moment $x_i$; $\delta_i$ is a uniformly distributed random number from 0 to 1; $\varepsilon$ is the one-sided limit of the measurement error in fractions of a unit (0<$\varepsilon$<1.0).

Given that the time moments of discrete measurements can be unevenly distributed, in the simulation the step between neighboring values of the argument was taken as a randomly distributed value:

$$h_i = h_m(1 + 2(\rho_i - 0{,}5)\Delta) \tag{14}$$

where $h_i$ is the step between two neighboring values of the argument (*h*-step), i.e. $h_i = x_i - x_{i-1}$; $h_m$ is the base (average) value of *h*-step; $\rho_i$ is a uniformly distributed random number from 0 to 1; $\Delta$ is the one-sided limit of relative variation of *h*-step in fractions of one (0<$\Delta$<1.0).

**Fig. 1** shows an example of one implementation ("run") of a simulation model of accelerated steel plate cooling, which hypothetically obeys function (11). The change in temperature over time is shown in **Fig. 1a**, and the change in cooling rate over time is shown in **Fig. 1b**. The values of parameters $a_1$, $a_2$, $\varepsilon$, $h_m$ and $\Delta$ are given in the caption to **Fig. 1**.

The degree of correspondence between the finite-difference and theoretical derivative was estimated by the value of their partial relative deviation in each *i*-th point:

$$r_i = \frac{y'_i - y'(x_i)}{y'(x_i)} \cdot 100\% \tag{15}$$

Using the values of $r_i$ obtained in one run of the simulation model for all $i$, we calculated the frequency of $r_i$ falling into a particular $\pm R\%$ symmetric pocket (e.g., $\pm 25\%$ or $\pm 50\%$, etc.) in a given run. By averaging these frequencies over a large number ($N$) of runs, we calculated the statistical probability $P_R$ that the finite-difference derivative (9) would deviate from the theoretical derivative (12) by no more than a given value $\pm R\%$. Below the parameter $R$ will be called as *a deviation field*, and $P_R$ – as *the probability of not exceeding the deviation field R*.

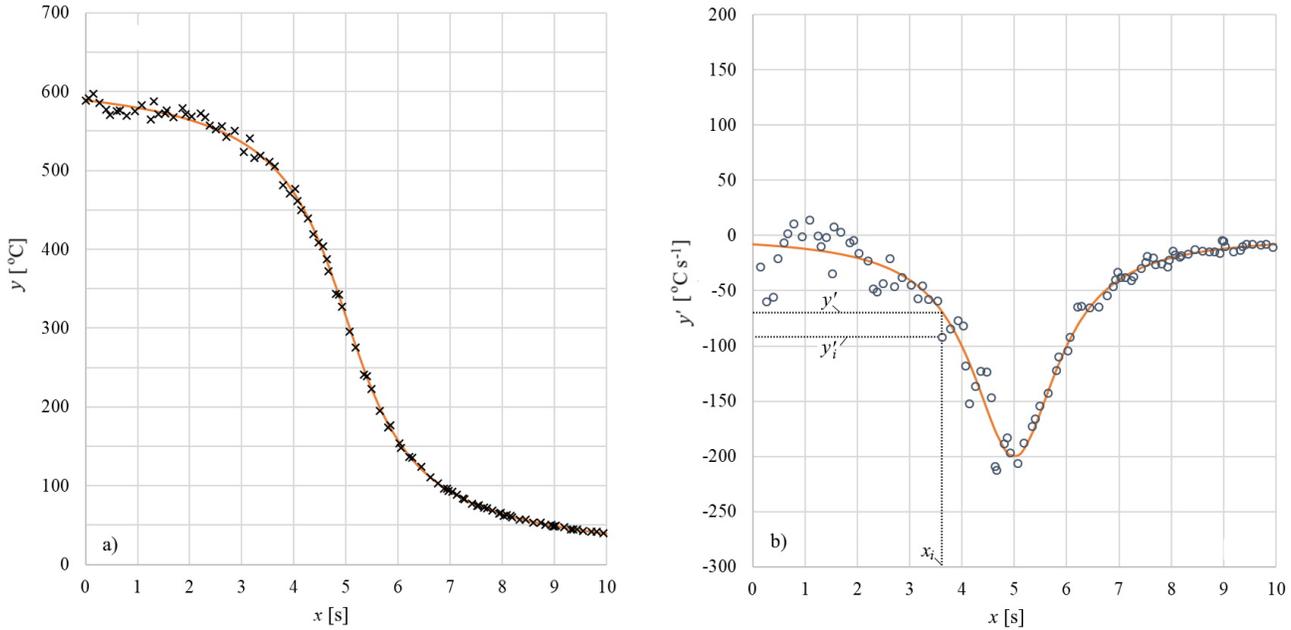

**Fig. 1.** Example of one realization ("run") of a simulation model of accelerated cooling of a steel plate obeying function (11) with parameters $a_1$=200 °C and $a_2$=5 s: a) - temperature data: solid line - graph of function $y(x)$ according to (11), dots - hypothetical empirical data $y_i$ calculated by (13) with $\varepsilon$=0.025, i.e. obtained with random deviations within $\pm 2.5\%$ from graph $y(x)$, in unequal time points $x_i$ with base step $h_m$=0.1 s and limit of relative variation from it $\Delta$=0.75; b) - first derivative of temperature on time: solid line - graph of function $y'(x)$ according to (12), points - evaluation by formula (9) at $n$=0.75.

This is explained in **Fig. 2**, which shows the sampling distribution function of the value $r_i$ and the probability $P_R$ for the example shown in **Fig. 1**. The statistical characteristics shown in **Fig. 2** are obtained with three different values of the weighting factor in formula (9): $n = 0$, $n = 0.75$, and $n = 4$. For example, if the finite-difference derivative by this formula is calculated at $n = 0.75$, then 81% of all $r_i$ values are less than (+25)% and 19% of all $r_i$ values are less than (-25)% (see **Fig. 2a**). Accordingly, 62% of all $r_i$ values fall into the region of $\pm 25\%$, that is, the probability of not exceeding the deviation field of $\pm 25\%$ is $P_{25}$=62% (see **Fig. 2b**). At other values of the weighting factor the security of $P_{25}$ has a smaller value: $P_{25}$=60% and $P_{25}$=52% at $n = 0$ and $n = 4$, respectively.

Similar calculations were performed for other test functions and parameters $\varepsilon$, $h_m$, and $\Delta$ in formulas (13)-(14). Some results are shown in **Fig. 3**, where the dependence of $P_R$ on the weighting coefficient $n$ is presented. The solid lines refer to function (11), and the dotted lines to another test function (**Fig. 4**):

$$y(x) = c_1 x^3 + c_2 x \cdot \sin(c_3 x) \tag{16}$$

with a rapidly changing derivative:

$$y'(x) \equiv \frac{dy(x)}{dx} = 3c_1 x^2 + c_2 c_3 x \cdot \cos(c_3 x) + c_2 \cdot \sin(c_3 x) \tag{17}$$

where $c_1$, $c_2$ and $c_3$ are constant parameters.

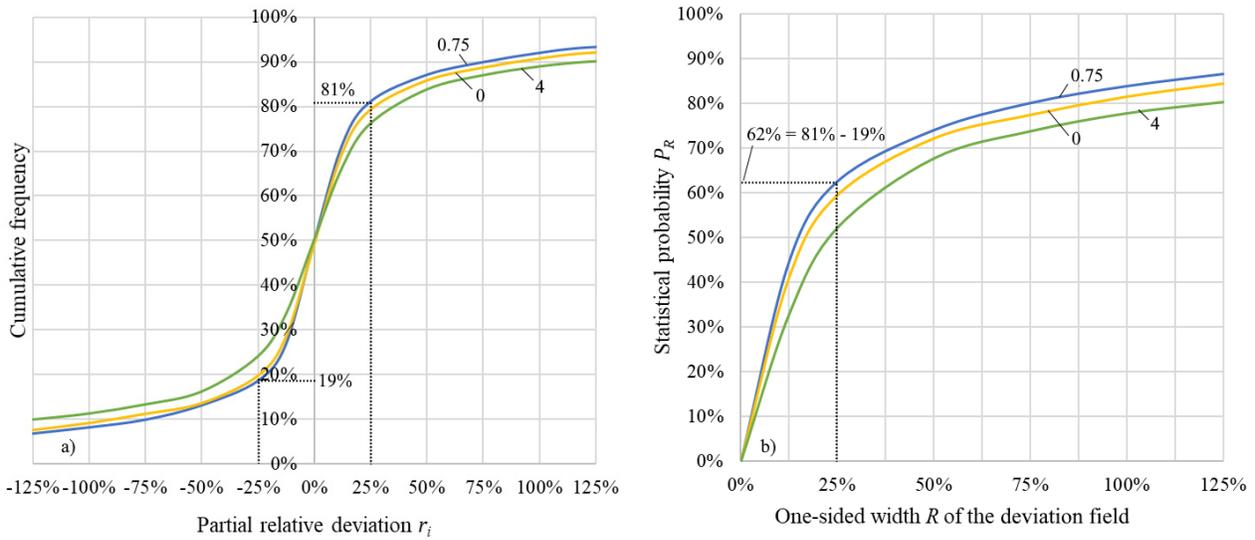

**Fig. 2.** Statistical characteristics of the distribution of the value of partial relative deviation under the conditions of the example shown in **Fig. 1** in a sample of 50 runs ($N = 50$): *a* - the distribution function of the private relative deviation $r_i$; *b* - the statistical probability $P_R$ that this deviation $r_i$ falls into a certain symmetric pocket $\pm R\%$ (the probability of not exceeding the deviation field $R$). The numbers at the curves denote the value of the weight coefficient n in formula (9).

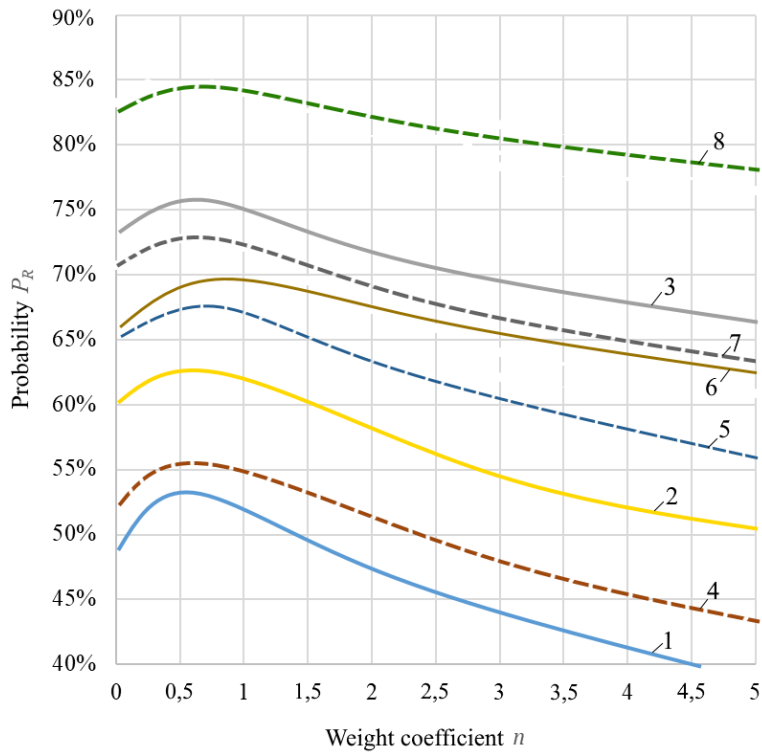

**Fig. 3.** Dependence of $P_R$ probability on weight coefficient $n$, obtained by simulation of processes obeying function (11) with parameters $a_1=200$ and $a_2=5$ (curves 1-3) and function (16) with $c_1=0,05$, $c_2=4$ and $c_3=1.5$ (curves 4-8). The curves refer to different values of ultimate measurement error $\varepsilon$ and sampling parameters in formulas (13)-(14), as well as the deviation field width $R$: 1 - $\varepsilon = 0,075$, $h_m = 0.1$, $\Delta = 0.75$; $R= \pm 50\%$; 2 - $\varepsilon = 0.025$, $h_m = 0.1$, $\Delta = 0.75$; $R= \pm 25\%$; 3 - $\varepsilon = 0.025$, $h_m = 0.2$, $\Delta = 0.75$; $R= \pm 25\%$; 4 - $\varepsilon = 0.05$, $h_m = 0.15$, $\Delta = 0.75$; $R= \pm 25\%$; 5 - $\varepsilon = 0.075$, $h_m = 0.2$, $\Delta = 0.75$; $R= \pm 50\%$; 6 - $\varepsilon = 0.025$, $h_m = 0.2$, $\Delta = 0.75$; $R= \pm 25\%$; 7 - $\varepsilon = 0.05$, $h_m = 0.15$, $\Delta = 0.75$; $R= \pm 50\%$; 8 - $\varepsilon = 0.05$, $h_m = 0.2$, $\Delta = 0.75$; $R= \pm 50\%$.

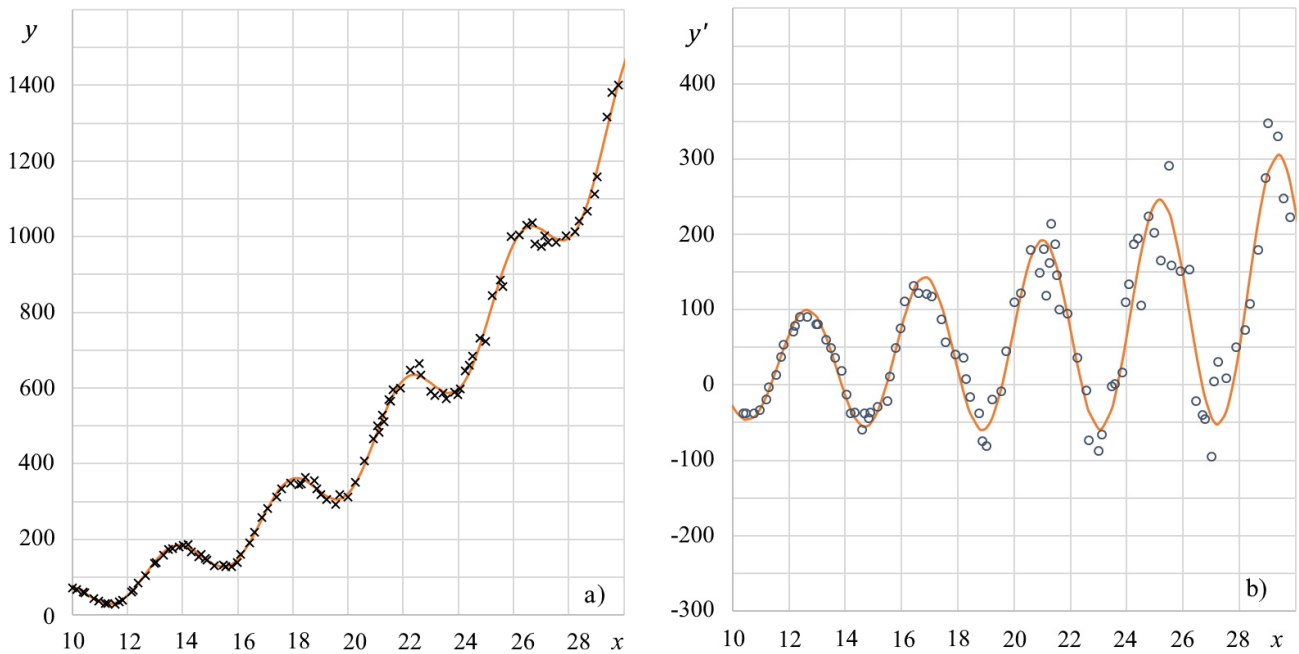

**Fig. 4**. An example of one realization ("run") of the simulation of the process obeying function (16) with parameters $c_1$=0.05, $c_2$=4 and $c_3$=1.5: a) - data on the dependent variable $y$: the solid line is the graph of the function $y(x)$ according to (16), the points are the hypothetical empirical values of $y_i$ calculated by the formula (13) with $\varepsilon$=0.05, i.e. obtained with random deviations within ±5% from graph $y(x)$, in unequally spaced points by time $x_i$ with the basic step $h_m$=0.2 and the limit of its relative variation $\Delta$=0.75; b) - first derivative of function: solid line - graph of function $y'(x)$ according to (17), points - finite-difference estimates of first derivative by formula (9) at $n$=0.75.

## RESULTS

The obtained data show that if the limit of the random error of function measurements is ±2% or more (i.e., $\varepsilon \geq 2\%$) for completely different functions and different sampling parameters ($h_m$, $\Delta$) the optimal values of the weight coefficient $n$ are in a narrow range: from about 0.5 to 0.75. At values of $n$ within this range, the best correspondence between the finite-difference values according to formula (9) and the theoretical values of the first derivative is achieved. Here we note that when $n = 0.5$ and the step between the points is equal, i.e., when $h_i$=const$\equiv h$, formula (9) becomes identical to the expression given in [13, p. 328].

Taking $n$=2/3 in formula (9) as a value in the middle of the optimal range of 0.5...0.75, we obtained the final form of a simple engineering formula for approximating the first derivative when analyzing unequally distributed empirical data with errors:

$$(y'_i)_{5uS} = \frac{3y_{i+2} + 2y_{i+1} - 2y_{i-1} - 3y_{i-2}}{3x_{i+2} + 2x_{i+1} - 2x_{i-1} - 3x_{i-2}} \qquad (18)$$

**Fig. 5** shows the results of calculations by formula (18) in comparison with the results by formulas (1) and (3). It can be seen that the proposed engineering formula gives significantly more accurate results for the smallest measurement errors of the function due to the fact that it simultaneously takes into account the smoothing of empirical data. For example, according to **Fig. 5a**, if the measurement error limit of the function (11) is ±2,5 %, then at estimation of the first derivative by the formula (18) the deviation field security ±25 % will be above 60 %, while at estimation by the formulas (1) and (3) - below 30 %.

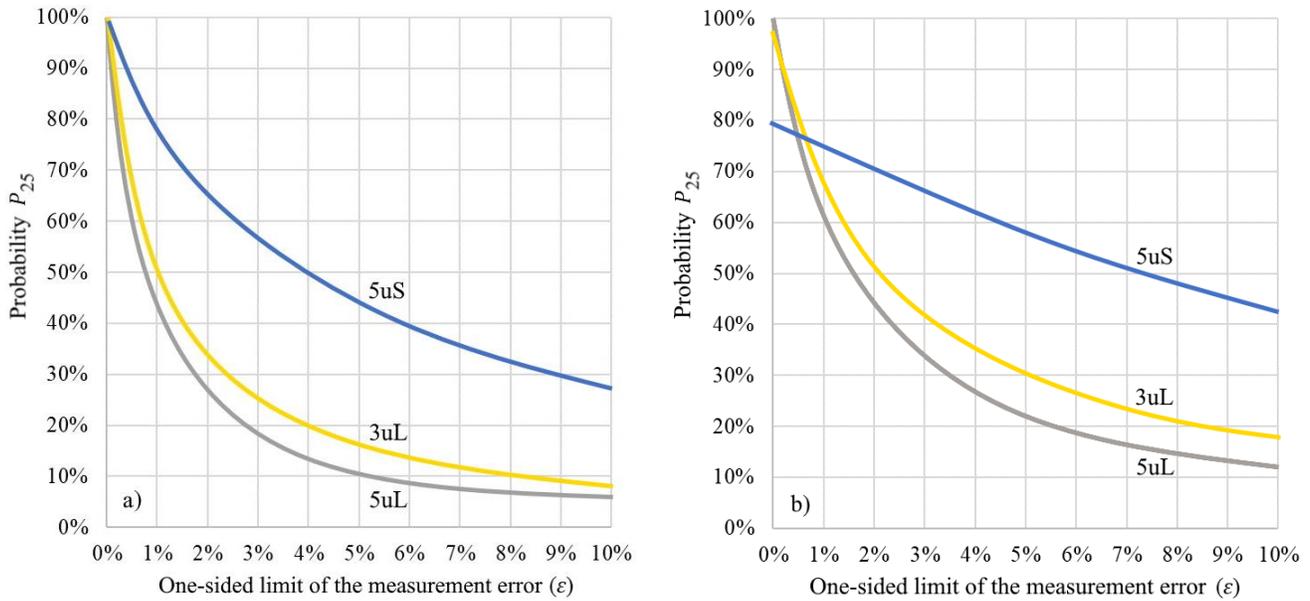

**Fig. 5.** Provision of the field of deviations ±25% from the true values of the first derivative unknown function in cases of calculation from empirical data with errors: 3*uL* by formula (1), 5*uL* by formula (3), 5*uS* by the authors formula (18): *a* - simulation on test function (11) with parameters $a_1$=200; $a_2$=5; $h_m$=0.1; $\Delta$=0.75; *b* - simulation on test function (16) with parameters $c_1$=0.05; $c_2$=4; $c_3$=1.5; $h_m$=0.2; $\Delta$=0.75.

## CONCLUSION

In a wide variety of technical problems, the use of known formulas for estimating the derivative of an empirical function from its values measured at discrete points leads to errors. This is explained by the fact that such formulas either take into account unequal intervals between measurement points, but do not take into account the error of measurements themselves, or vice versa, they take into account the possible error of measurements, but do not take into account the unequal location of measurement points.

The authors have proposed a heuristic formula (18), which simultaneously takes into account both factors inherent in any tabulated empirical function: 1) unequally spaced values of the argument and 2) the error in the measured values.

The results of simulation on various test functions show that even with a small random error of measurements (from ±1%) the formula (18) allows us to drastically improve the accuracy of the estimation of the first derivative of the empirical function.